\documentclass{ws-ijmpa}

\begin{document}

\markboth{R. Delbourgo}
{Grassmannian duality and the particle spectrum}
\catchline{}{}{}{}{}

\title{Grassmannian Duality and the Particle Spectrum}

\author{Robert Delbourgo}

\address{School of Physical Sciences, University of Tasmania, Locked Bag 37 GPO\\ Hobart, Tasmania 7001,AUSTRALIA 7001\\
bob.delbourgo@utas.edu.au}

\maketitle
 
%\begin{history}
%\received{Day Month Year}
%\revised{Day Month Year}
%\end{history}
        
\begin{abstract}
Schemes based on anticommuting scalar coordinates, corresponding to properties,
lead to generations of particles naturally. The application of Grassmannian duality
cuts down the number of states substantially and is vital for constructing
sensible Lagrangians anyhow. We apply duality to all of the subgroups within the
{\em classification} group SU(3)$\times$SU(2)$_L\times$SU(2)$_R$, which encompasses 
the standard model gauge group, and thereby determine the full state inventory; this includes
the definite prediction of quarks with charge -4/3 and other exotic states. Assuming
universal gravitational coupling to the gauge fields and parity even
property curvature, we also obtain $4\sin^2\theta_w = 1 - 2\alpha/3\alpha_s$
which is not far from the experimental value around the $M_Z$ mass.

\end{abstract}
\ccode{PACS: 11.10Kk,11.30.Hv,11.30.Pb,12.10.-g}

\section{Generations and Duality} %Section I

The existence of generations of particle families is one of the deepest conundrums
in theoretical physics. Who can forget Rabi's exclamation upon the discovery of the
muon: "Who ordered that?". The invocation of a new family symmetry group offers no
explanation to Rabi but can serve as a useful classification tool and it can even be gauged.
Over the last few years we have suggested that the full description of events through
a spacetime $x$ enlarged by anticommuting complex scalar coordinates $\zeta$, which
correspond to fundamental properties, can serve to explain naturally the repetition
of particle families. Put another way, we have proposed that the attributes
of events represent the extra dimensions which everyone is chasing; such an approach
is distinct from other attempts at unification. We have gone on to show that the forces and their 
parity violation (at low energies anyway) can be accommodated through the curvature of 
the metric in the enlarged spacetime $X=(x,\zeta,\bar{\zeta})$ particularly via the  
$x-\zeta$ sector, making a distinction between left and right attributes.

The article\cite{RD} on ``The Force and Gravity of Events''  contains a detailed reference list
describing progressive steps made in researching this scenario.
Our investigations have shown that a bare minimum of three (parity blind) colour
properties $\zeta^1,\zeta^2, \zeta^3$ plus two electroweak properties $\zeta^0,\zeta^4$,
{\em further distinguished by their parities}, 
can potentially produce the known generations and can even describe the standard model 
(SM) interactions correctly; the force fields arise through the gauge mesons
embedded in the frame vectors that connect spacetime to property at the same time that
gravitational properties in the spacetime sector stay impartial with respect to handedness,
at least semiclassically. On several occasions\cite{RD} we have alluded to some
of the multiplets engendered by such a scheme, but we have never provided a {\em full}
inventory of the states which ensue. That is the purpose of this article  as well as 
constructing the interactions between all the fields posited by the standard model.

The state zoo predicted by such a scheme arises when the superfields $\Phi(X)~\&~\Psi(X)$
are expanded in $\zeta$ and its conjugate $\bar{\zeta}$. There
one encounters a {\em finite} number of terms $(\bar{\zeta})^r(\zeta)^s\phi_{(r,s)}$ \&
$(\bar{\zeta})^r(\zeta)^s\psi_{(r,s)}$, where $r,s$ run up to the number of properties,
with due care taken to obey the spin-statistics theorem. Even so, a considerable number 
of expansion terms are produced even with only $N=5$ properties.
However once we realise that unitary representations $\phi,\psi$ are unaffected by the 
`duality' substitution $(r,s)\rightarrow (N-s,N-r)$, we can apply a duality symmetry to
reduce substantially the number of terms in the superfield expansions. Interestingly,
the imposition of duality is essential to obtain sensible Lagrangians for the component
fields after integrating over property. This is explained at length in the next section
as well as the appendix. We make particular choices of duality for each of the subgroups 
of the SM and, following that, enumerate the full list of right and left states
in section 3, paying particular heed to the many chargeless Higgs fields held in $\Phi$
since they are responsible for mass generation of bosons and fermions through the Yukawa
interaction. The use of a universal property curvature polynomial leads to a determination 
of the weak mixing angle by demanding universal gravitational coupling to the gauge fields;
the value of that angle is close to experimental measurements before taking account of
quantum corrections. We conclude with some remarks about this whole approach 
in the final section.

\section{The Need for Duality: Examples} %Section II

\subsection{One property}
We start with just a single property, corresponding to one complex $\zeta$,
representing electricity. Paying no regard to parity at this stage and 
reminding ourselves that bosons are associated with even powers and 
fermions with odd powers, the expansions of the superfields in $\zeta$
become rather trivial:
\begin{eqnarray}
\Phi(x,\zeta,\bar{\zeta}) &=& A(x) + \bar{\zeta}\zeta B(x)\nonumber \\
\Psi(x,\zeta,\bar{\zeta}) &=& \bar{\zeta}\psi(x) + \psi^c(x)\zeta \nonumber \\
\overline{\Psi}(x,\zeta,\bar{\zeta})&=& -\overline{\psi}\zeta+\bar{\zeta}\overline{\psi^c}.
\end{eqnarray}
Note that $\Phi$ and $\Psi$ (the latter carries a spinorial index) are both bosonic but
their expansion coefficients have the correct spin-statistics connection.
We shall be integrating over $\zeta$ so we will use the convention\footnote{
When dealing with $N$ properties, consult the Appendix for our convention.} that
$\int (d\zeta d\bar{\zeta}) \,(\bar{\zeta}\zeta)= 1$. It follows that the integrated bilinear
\[
\int (d\zeta d\bar{\zeta}) \,\Phi\Phi = 2AB,
\]
can hardly serve as a suitable Lagrangian. This is where the concept of Grassmannian 
duality comes to the rescue. With only one property, dualization, indicated by 
the superscript $\times$, corresponds to
\begin{equation}
1^\times = \bar{\zeta}\zeta, \quad  (\bar{\zeta}\zeta)^\times = 1, \quad
\zeta^\times = \zeta, \quad (\bar{\zeta})^\times = \bar{\zeta}.
\end{equation}
Therefore the selfdual $\zeta$ polynomials are $(1 + \bar{\zeta}\zeta)$ and $\zeta$,
while the anti-selfdual polynomial is $(1 - \bar{\zeta}\zeta)$ with $\zeta\rightarrow 0$.
Requiring selfduality means setting $A=B$ and permits suitable Lagrangian
bilinears:
\[
\int (d\zeta d\bar{\zeta}) \,\Phi\Phi = 2A^2,\qquad 
\int (d\zeta d\bar{\zeta}) \,\overline{\Psi}\Psi = \bar{\psi}\psi+\overline{\psi^c}\psi^c
= 2\bar{\psi}\psi.
\]
Double dualization amounts to the identity operation, and self-evidently the total number of 
component fields for anti-selfdual plus selfdual adds up to what one would get without
imposing those conditions.

\subsection{Two properties}
Our starting point is the pair of properties, $\zeta^0$ signifying neutrinicity and 
$\zeta^4$ signifying charge -1 leptonicity; for the present we disregard colour and therefore 
strong interactions (to which we presently assign the labels  $\zeta^{1,2,3}$, signifying colour down-type quarks). Later on we have to distinguish between left and right leptonicity.

Here we are dealing with an SU(2) subgroup of Sp(4). When we expand the superfields
in powers $(\bar{\zeta})^s(\zeta)^r$ with $r,s$ running from 0 to 2 we obtain
the square given in Table 1 where it will be seen that reflection about the
diagonal corresponds to charge conjugation, whereas reflection about the
cross-diagonal corresponds to taking the dual.

\begin{table}[th]
\tbl{Isospin states $(\bar{\zeta})^r(\zeta)^s$; $r+s$ odd for fermions, $r+s$ even for bosons.
Numerical entries indicate dimensions of representations.}
{\begin{tabular}{|c||c|c|c|}  
\hline
$s\backslash r$ & 0 & 1 & 2 \\ 
\hline
\hline
   0 &  1 & 2 &  1\\  
\hline
   1 & $\bar{2}$ & $1\oplus 3$ &  2\\ 
\hline
  2 & $\bar{1}$ &  $\bar{2}$ &  1   \\
\hline
\end{tabular} \label{ta1}}
\end{table}  

With regard to duality signs, we use the antisymmetric Levi-Civita symbols, 
$\epsilon^{\bar{0}\bar{4}}=1, \epsilon^{40}=1$ with
$\epsilon^{\bar{\lambda}\bar{\mu}}\epsilon^{\mu\nu}=\delta^{\bar{\lambda}\nu}$.
Introduce the abbreviation $Z\equiv \zeta^{\bar{\mu}}\zeta^\mu\equiv \bar{\zeta}\zeta$, so 
$Z^2/2=\zeta^{\bar{4}}\zeta^4\zeta^{\bar{0}}\zeta^0$. Then the duality operation results in:
\begin{eqnarray}
1^\times &=& Z^2/2, \quad (\zeta^\mu)^\times = Z\zeta^\mu,\quad 
(\zeta^0\zeta^4)^\times = -(\zeta^0\zeta^4)^\times \nonumber \\
(\zeta^{\bar{0}}\zeta^0)^\times &=& -\zeta^{\bar{4}}\zeta^4, 
{\rm ~so~} Z^\times = -Z,\,\,{\rm~while~for~the~triplet},\nonumber \\
(\zeta^{\bar{0}}\zeta^4)^\times &=& \zeta^{\bar{0}}\zeta^4,\quad
(\zeta^{\bar{0}}\zeta^0 - \zeta^{\bar{4}}\zeta^4)^\times =
(\zeta^{\bar{0}}\zeta^0 - \zeta^{\bar{4}}\zeta^4),\quad 
(\zeta^{\bar{4}}\zeta^0)^\times = \zeta^{\bar{4}}\zeta^0,
\end{eqnarray}
plus all their adjoints. This then allows us to separate the two possibilities\footnote{
In a previous paper we differed from the convention here in that our integration measure 
had the opposite sign. Thus what we regarded as anti-selfdual then becomes selfdual now.}:

\subsubsection{Selfdual expansion}
For fermions we get
\begin{eqnarray}
\sqrt{2}\,\Psi &=& \zeta^{\bar{0}}(1+Z)\nu +
               \zeta^{\bar{4}}(1+Z)\ell,\nonumber \\
 -\sqrt{2}\,\overline{\Psi} &\equiv& \bar{\nu}(1+Z)\zeta^0 +
               \bar{\ell}(1+Z)\zeta^4,
\end{eqnarray}
 ignoring~conjugates which only serve to double the final results. 
 The selfdual Bose superfield expansion on the other hand reads
\begin{equation}
\sqrt{2} \Phi =  \phi(1 + Z^2/2) + \zeta^{\bar{\mu}}\rho^{\mu\bar{\nu}}\zeta^\nu
 \end{equation}
 where $\phi$ is a singlet and $\rho$ is an isotriplet. These fields comprise eight 
 components in toto.
 
\subsubsection{Anti-selfdual expansion}

Here we get instead
\begin{eqnarray}
\sqrt{2}\,\Psi &=& \zeta^{\bar{0}}(1-Z)\nu +
               \zeta^{\bar{4}}(1-Z)\ell,\nonumber \\
\sqrt{2}\,\overline{\Psi} &\equiv& \bar{\nu}(1-Z)\zeta^0 +
               \bar{\ell}(1-Z)\zeta^4,
\end{eqnarray}
plus conjugates, and
\begin{equation}
 \Phi =  \phi(1 - Z^2/2) + \sigma Z + \zeta^{\bar{4}}\zeta^{\bar{0}}\varphi
 + \overline{\varphi}\zeta^0\zeta^4,
 \end{equation}
where $\varphi$ is a complex isosinglet. This set also contains eight components. 
Added to the previous (selfdual) set we get a total of 16 states, agreeing with the
number of coefficients expected for these expansions without imposing duality.
 
\subsubsection{Right and Left}

Presently we will be distinguishing between left and right properties $\zeta_L, \zeta_R$ as
demanded by the gauge interactions of the standard model; so the above analysis can be
applied to each of the SU(2) {\em classification} subgroups in the overarching group 
$SU(2)_R\times SU(2)_L$, though specifically right-handed gauge fields do not 
exist in the standard model.  

Thus the allowed {\em selfdual} combinations with their dimensionalities, for each chirality, are
\begin{eqnarray}
{\rm Bosons~}B^D:-&&\,\,\mathbf{1}:(1+Z^2/2),\quad\mathbf{3}: \bar{\zeta}\zeta\nonumber \\
\&\quad{\rm Fermions~}F^D:-&&\,\,\bar{\mathbf{2}}:\bar{\zeta}(1+Z),\quad\mathbf{2}:\zeta(1+Z),
\end{eqnarray}
with the singlet $\zeta^0\zeta^4$ eliminated. With reference to the triplet, it has the three components $(\zeta^{\bar{4}}\zeta^0, [\zeta^{\bar{4}}\zeta^4\!-\! \zeta^{\bar{0}}\zeta^0]/\sqrt{2},
-\zeta^{\bar{0}}\zeta^4) {\rm ~carrying~ charges~}(1,0,-1).$
It is to be understood that this shorthand for the triplet, viz. $\bar{\zeta}\zeta$, will appear
subsequently for each of the  left or right SU(2) subgroups.

On the other hand, the {\em anti-selfdual} combinations with their dimensionalities,
for each chirality, are
\begin{eqnarray}
{\rm Bosons~} B^A:-&&\,\,\mathbf{1}:(1-Z^2/2),\,\,Z,\,\,\zeta^0\zeta^4,\,\,
\zeta^{\bar{4}}\zeta^{\bar{0}}  
\nonumber\\
\&\quad{\rm Fermions~}F^A:-&&\,\,\bar{\mathbf{2}}:\bar{\zeta}(1-Z),\quad\mathbf{2}:\zeta(1-Z).
\end{eqnarray}

\subsection{Three properties}

We now turn to the strong interaction sector and colour to see what selfduality has to say
about this. Table 2 is the analogue of Table 1, before imposition of duality, and contains the 
following entries for the SU(3) representations arising from $\zeta,\bar{\zeta}$ expansion.

\begin{table}[th]
\tbl{The colour multiplets $(\bar{\zeta})^s(\zeta)^r$; $r+s$ odd for fermions, even for bosons.
Numerical entries indicate dimensions of representations.}
{\begin{tabular}{|c||c|c|c|c|}  
\hline
$s\backslash r$ & 0 & 1 & 2 & 3\\ 
\hline
\hline
   0 &  1 & 3 & $\bar{3}$ &  1\\  
\hline
   1 & $\bar{3}$ & $1\oplus\bar{8}$ & $3\oplus \bar{6}$ & $\bar{3}$ \\ 
\hline
  2 & 3 & $\bar{3}\oplus 6$ &  $1\oplus 8 $ &  3   \\
\hline
 3 & 1 & 3 & $\bar{3}$ & 1 \\
 \hline
\end{tabular} \label{ta2}}
\end{table} 
 Observe that the $\bar{6}$-fold colour multiplet contains such components as
$\zeta^1\zeta^{\bar{2}}\zeta^3,\zeta^1(\zeta^{\bar{2}}\zeta^2 -\zeta^{\bar{3}}\zeta^3) $,
in contrast to the triplet $\zeta^1(\zeta^{\bar{2}}\zeta^2+\zeta^{\bar{3}}\zeta^3)=\zeta^1 Z$,
where $Z\equiv (\zeta^{\bar{1}}\zeta^1+\zeta^{\bar{2}}\zeta^2+
\zeta^{\bar{3}}\zeta^3)$. The duality signs are encapsulated by the typical entries:
\begin{eqnarray}
1^\times&=&Z^3/6, \quad Z^\times =Z^2/2,\quad 
\zeta^{\bar{2}}\zeta^1=-Z\zeta^{\bar{2}}\zeta^1~({\rm octet})\nonumber \\
(\zeta^1\zeta^2)^\times&=&Z(\zeta^1\zeta^2),\quad 
(\zeta^1\zeta^2\zeta^3)^\times=\zeta^1\zeta^2\zeta^3,\quad 
(\zeta^1Z)^\times =\zeta^1Z\nonumber\\
(\zeta^1\zeta^{\bar{2}}\zeta^3)^\times&=&\zeta^1\zeta^{\bar{2}}\zeta^3, \quad
(\zeta^1(\zeta^{\bar{2}}\zeta^2 -\zeta^{\bar{3}}\zeta^3))^\times
= \zeta^1(\zeta^{\bar{2}}\zeta^2 -\zeta^{\bar{3}}\zeta^3)~{\rm (sextet)}.
\end{eqnarray}
The consequence is that the {\em selfdual} combinations for 3 properties 
(including conjugates), with their SU(3) dimensionalities comprise the sets:
\begin{eqnarray}
{\rm Bosons~}B^D:-&&\quad\mathbf{1}:(1+Z^3/6),\,\,Z(1+Z/2),\nonumber \\
 & &\quad\bar{\mathbf{3}}:\zeta\zeta(1+Z),\,\,\,\mathbf{3}:\bar{\zeta}\bar{\zeta}(1+Z),
\,\,\,\mathbf{8}:\zeta\bar{\zeta}(1-Z);\\
\&\quad{\rm Fermions~}F^D:-&&\quad\mathbf{3}:\zeta(1+Z^2/2),\,\,
\,\bar{\mathbf{3}}:\bar{\zeta}(1+Z^2/2),\nonumber \\
&&\quad \bar{\mathbf{6}}:\zeta\zeta\bar{\zeta},\quad
\mathbf{6}:\zeta\bar{\zeta}\bar{\zeta},
\quad\mathbf{1}:\zeta\zeta\zeta,\,\bar{\zeta}\bar{\zeta}\bar{\zeta},
\end{eqnarray}
with the triplet $\zeta Z$ eliminated. This makes for a total of 36 selfdual states.
The {\em anti-selfdual states} number 28 states:
\begin{eqnarray}
{\rm Bosons~}B^A:-&&\quad\mathbf{1}:(1-Z^3/6),\,\,Z(1-Z/2),\nonumber \\
 & &\quad\bar{\mathbf{3}}:\zeta\zeta(1-Z),\,\,\,\mathbf{3}:\bar{\zeta}\bar{\zeta}(1-Z),
\,\,\,\mathbf{8}:\zeta\bar{\zeta}(1+Z);\\
\&\quad{\rm Fermions~}F^A:-&&\quad\mathbf{3}:\zeta(1-Z^2/2),\,\,\zeta Z\nonumber\\
&&\quad \bar{\mathbf{3}}:\bar{\zeta}(1-Z^2/2),\,\,\,\bar{\zeta}Z,
\end{eqnarray}
with the singlet $\zeta\zeta\zeta$ excised. Added to the selfdual 
ones, we get a total of 64 states, as expected. (For $N$ properties there are $4^N$ states
before constraining them by duality.)

Subsequently, when constructing Lagrangians, it is important to point out that the integrals
over property of quadratic combinations such as $B^AB^D~\&~\bar{F}^DF^A$ vanish identically.
However cubic terms, which arise when considering Yukawa couplings,
such as $B^AB^AB^D~ \&~ B^A\bar{F}^AF^D$,  survive property integration. That applies to
each of the subgroups, being independent of the number of properties.

\section{The full particle inventory}

At this point we can combine all of the attributes and determine the complete set of
states, taking into account parity too -- we did not do this previously when we just
considered the grand unified group SU(5) duality constraints. In order to decide what
type of duality is to be applied to each of the standard model subgroups, 
we must make sure that certain `unwanted' combinations are duly expunged and that 
we cover the known generations of quarks and leptons, as well as incorporating proper 
Higgs-like states. Another important consideration is that the fermionic states have to 
be constrained to be of two types, purely left or purely right; we cannot entertain 
a schizophrenic or ambidextrous possibility of mixing these two sorts of chirality 
{\em unless one of them is an ineffective singlet} in the product. On the contrary, the 
scalar fields {\em must bridge right and left}, which may dictate a different choice of duality 
from the fermions.

So let us start off with the quarks: colour triplets and weak isodoublets. Let 
$\zeta_c$ refer to chromic property (which strictly speaking should carry 
contravariant labels 1,2,3).  When we paid no heed to chirality and ignored duality 
we were able to identify two typical $(U,D)$ cases:
$(\bar{\zeta}_c\bar{\zeta}_c\zeta^0,\bar{\zeta}_c\bar{\zeta}_c\zeta^4)$ and
$(\zeta_c\zeta^{\bar{4}}\zeta^0,\zeta_c\zeta^{\bar{4}}\zeta^4{\rm~or~}
\zeta_c\zeta^{\bar{0}}\zeta^0)$
as having the correct colour and charge quantum numbers. Chromic duality will double these 
pairs (whether we pick colour dual or antidual) and with the first choice we can readily impose 
left or right-handed characteristics on the leptonic component. The second choice is more problematic because it involves a lepton-antilepton product. These must both be left {\em or}
right according to our criterion, so referring to Sec.2.2.3 and (8) we have to take
the {\em triplet} combination; this dictates selfduality for left. The right products 
lepton-antilepton accompanying them must not jeopardize the chiral purity and must
therefore remain right singlets so, as in (9), we have to pick anti-selfduality for right.
When we reverse overall chirality, similar sorts of selections must prevail.

Turning to the scalar bosons, we note that $\zeta^1\zeta^2\zeta^3\zeta^{\bar{4}}$ has the potential to serve as a Higgs boson. However it is apparent that a similar combination, 
$\zeta^1\zeta^2\zeta^3\zeta^4$, is a state with $F=2, Q=-2$ and is undesirable.
Besides, such a state would be either left or right and could not bridge chirality.
Noting that the colour singlet $\zeta^1\zeta^2\zeta^3$ disappears when
we adopt colour anti-selfduality this then indicates chromic anti-selfduality as being
the proper fermionic choice. We shall be guided by these considerations below.

Let $F^A_c, B^A_c$, $F^A_L,B^A_L$, $F^A_R,B^A_R$ signify the fermionic and 
bosonic parts of the combinations complying with antiduality, while $F^D_c,B^D_c$ etc.
signify those combinations which are selfdual. Then the full fundamental fermionic elements 
consist of the products $F_c^AF^A_RF^D_L,  F_c^AB_R^AB_L^D, B_c^AF^A_RB^D_L,
B_c^AB_R^AF_L^D$, in addition to the corresponding terms where one interchanges $L$ and 
$R$. The resulting buildup is based upon stating the attributes of the fundamental properties
determined by the action of the standard gauge fields, held in the spacetime-property
sector of the metric. These property attributes are listed in Table 3.
\begin{table}[th]
\tbl{The quantum numbers for weak left isospin, hypercharge and charge of the basic
properties. The conjugates $\bar{\zeta}$ naturally have the opposite quantum numbers.}
{\begin{tabular}{|c||c|c|c|}   \hline
Property & $T_{3L}$ & $Y$ & $Q$\\ 
\hline
\hline
   $\zeta^0_L$ & 1/2 & -1 & 0 \\  
\hline   
  $\zeta^4_L$ & -1/2 & -1 & -1 \\
\hline
  $\zeta^0_R$ & 0 & 0 & 0 \\ 
\hline
 $\zeta^4_R$ & 0 & -2 & -1 \\
 \hline 
 $\zeta^{1,2,3}$ & 0 & -2/3 & -1/3 \\
 \hline
\end{tabular} \label{ta3}}
\end{table}
The state inventory comes about through combinations of these properties. It is 
noteworthy that sums of these basic attributes correspond exactly to the quantum
number assignments made in the SM. Before compiling all these products
systematically, it pays to identify where the 
usual up-, down- quarks and leptons reside, ignoring duality and handedness for now
and focussing on their charges. We have already identified two triplet colour
combinations that serve that purpose, namely $(\bar{\zeta}_c\bar{\zeta}_c\zeta^0,
\bar{\zeta}_c\bar{\zeta}_c\zeta^4)$ and ($\zeta_c\zeta^{\bar{4}}\zeta^0,
\zeta_c\zeta^{\bar{4}}\zeta^4/\zeta_c\zeta^{\bar{0}}\zeta^0$). This highlights 
the point that {\em we should expect a new colour triplet quark of charge} -4/3, identified 
as the combination $\zeta_c\zeta^{\bar{0}}\zeta^4$., which is part
of a weak isotriplet. Turning to the leptons, ${\cal L}\equiv(\nu, \ell$) 
are of course based on dualized generations of fundamental properties ($\zeta^0,\zeta^4$).
where it should be noted from (8) that $\zeta_L^0\zeta_L^4$ and $\zeta_R^0\zeta_R^4$ 
vanish for a selfdual choice. 

\subsection{The Fermionic States}

As explained above, possible combinations to be considered are multiplicative sets
$F_c^AB_R^DB_L^A, F_c^AB_R^AB_L^D, B_c^AB_R^DF_L^A, B_c^AF_R^AB_L^D$.
These are best summarised by listing the direct product representations and the dimensions
of the 3 {\em classification} groups: $SU(3)_c\otimes SU(2)_R\otimes SU(2)_L$;  
their associated properties are stated in eqs. (8),(9),(11) and (12).
In this way we obtain the left fermion combinations (including conjugates)
\begin{eqnarray}
F_c^A B_R^A B_L^D\supset&&\psi_L\sim \{3,3,\bar{3},\bar{3}\} \otimes \{1,1\}\otimes\{1,3\}\\
B_c^A B_R^A F_L^D\supset&&\quad\quad \{1,1,\bar{3},3,8\}\otimes\{1,1\}\otimes\{\bar{2},2\}
\end{eqnarray}
while the right combinations are
\begin{eqnarray}
F_c^A B_R^D B_L^A\supset&&\psi_R \sim \{3,3,\bar{3},\bar{3}\} \otimes \{1,3\}\otimes\{1,1\}\\
B_c^A F_R^D B_L^A\supset&&\quad\quad \{1,1,\bar{3},3,8\}\otimes\{2,\bar{2}\}\otimes\{1,1\}.
\end{eqnarray}
Identifying the quarks and paying little heed to the sizes of the SU(2)$_R$ 
representations since they are not gauged we can pick out the left quark products:
\[  
\{3,3\}\otimes \{1,1\} \otimes 3 : \zeta\{1-Z^2/2),Z\}\,.\,\{(1-Z_R^2/2), Z_R\}\,.\,
\zeta_L\bar{\zeta}_L
\]
\[
 3 \otimes\{1,1\}\otimes 2: \bar{\zeta}\bar{\zeta}(1-Z)\,.\,\{(1-Z_R^2/2),Z_R\}\,.\,
 \zeta_L(1+Z_L).
\]
with similar products where left and right are interchanged (plus their conjugates). 
The first and second generation 
of quarks correspond to weak isodoublets but the other quark generations occur in weak
isotriplets!  So far as leptons are concerned we have to seek colour singlets so must study
the set
\begin{eqnarray}
B_c^A B_R^A F_L^D\supset&&\psi_L\sim \{1,1\}\otimes\{1,1\}\otimes\{\bar{2},2\}\\
B_c^A F_R^D B_L^A\supset&&\psi_R\sim\{1,1\}\otimes\{\bar{2},2\}\otimes\{1,1\},
\end{eqnarray}
corresponding to the leptonic products
\[
\psi_L\sim \{(1-Z^3/6),Z(1-Z/2)\}\,.\,\{(1-Z_R^2/2),Z_R\}\,.\,\zeta_L(1-Z_L)
\]
\[
\psi_R\sim \{(1-Z^3/6),Z(1-Z/2)\}\,.\,\zeta_R(1-Z_R)\,.\,\{(1-Z_L^2/2),Z_L\}.
\]
We recognize {\em four} generations of leptons; so here is another prediction of this scheme.

\subsection{The Bosonic States}
Here one is obliged to take the following dual multiplicative combinations  so as to bridge 
left and right. Relying upon upon anti-selfdual combinations for the
chromic sector, and because $B_R^D B_L^A, B_R^A B_L^D$ does not lead to 
ambidextrous products, we might be tempted to consider
\begin{eqnarray*}
B_c^A F_R^A F_L^D:&&\phi\sim \{1,1,\bar{3},3, 8\}\otimes\{\bar{2},2\}\otimes\{\bar{2},2\}\\
B_c^A F_R^D F_L^A:&&\qquad\{1,1,\bar{3},3, 8\}\otimes\{\bar{2},2\}\otimes\{\bar{2},2\},
\end{eqnarray*}
with weak quartets contained within the colour singlet set:
\begin{eqnarray*}
B_c^A F_R^A F_L^D\supset \{(1-Z^3/6),Z(1-Z/2\}\,.\,\{\zeta_R,\bar{\zeta}_R\}(1-Z_R)\,.\,
     \{\zeta_L,\bar{\zeta}_L\}(1+Z_L)\\
B_c^A F_R^D F_L^A\supset \{(1-Z^3/6),Z(1-Z/2\}\,.\,\{\zeta_R,\bar{\zeta}_R\}(1+Z_R)\,.\,
     \{\zeta_L,\bar{\zeta}_L\}(1-Z_L).
\end{eqnarray*}
However it turns out that the above choices do {\em not} enjoy Yukawa interactions with
the quark and lepton generations (at least in flat space) because of property
integration rules. Instead for bosons we are led to relaxing the condition that
the chromic sector must be anti-selfdual and consider the following bosonic combinations 
which are {\em overall} selfdual and which {\em do} interact with the fermions:
\begin{eqnarray}
B_c^DF_R^D F_L^D:&&\phi\sim \{1,1,\bar{3},3, 8\}\otimes\{\bar{2},2\}\otimes\{\bar{2},2\}\\
B_c^D F_R^A F_L^A:&&\qquad\{1,1,\bar{3},3, 8\}\otimes\{\bar{2},2\}\otimes\{\bar{2},2\}.
\end{eqnarray}
Here the Higgs fields are contained within the colour singlet set,
\begin{eqnarray}
B_c^A F_R^A F_L^D\supset \{(1+Z^3/6),Z(1+Z/2\}\,.\,\{\zeta_R,\bar{\zeta}_R\}(1+Z_R)\,.\,
     \{\zeta_L,\bar{\zeta}_L\}(1+Z_L)\\
B_c^A F_R^D F_L^A\supset \{(1+Z^3/6),Z(1+Z/2\}\,.\,\{\zeta_R,\bar{\zeta}_R\}(1-Z_R)\,.\,
     \{\zeta_L,\bar{\zeta}_L\}(1-Z_L).
\end{eqnarray}
This last selection will become clearer when we consider gauge field and Yukawa interactions
where we also arrange that the superfield $\Phi$ is even under left-right interchange.

\section{Interactions}

As there appear to be a menagerie of exotic states (for which there is as yet 
no evidence) as well as quasi-conventional states, 
let us just focus on the standard  particles that are contained in amongst
the tables of bosons and fermions listed in the previous section.  We should remind
ourselves that the only gauged groups are QCD and the electroweak group, though
other researchers --- keen on left right symmetry --- might advocate more gauging. We need 
to check that this subset of states\footnote{We will include the quark family comprising 
charge -4/3, as that is one of the prominent predictions of the present scheme.} 
enjoy interactions expected of them through the metric
element connecting property to spacetime. But before launching into that we also need
to ascertain that the gauge field Lagrangians themselves emerge correctly from
the generalized scalar curvature of the extended space.

\subsection{Gauge field interactions}
We therefore turn to the extended metric which permits the following general form respecting
gauge invariance. Referring to Table 3,
\begin{eqnarray}
x-x~{\rm sector},~G_{mn}&=&g_{mn}C +``\,l^2\bar{\zeta}(A_mA_n+A_nA_m)\zeta C'/2\,",\\
x-\zeta_L~{\rm sector},~G_{m\zeta_L}&=&-il^2\bar{\zeta}_L(gW_m\!\cdot\!\tau-g'B_m) C'/4,\\
x-\zeta_R~{\rm sector},~G_{m\zeta_R}&=&-il^2\bar{\zeta}_Rg'B_m(\tau_3-1)/ C'/4\\
x-\zeta~{\rm sector},~G_{m\zeta}&=&-il^2\bar{\zeta}
    (fV_m\!\cdot\!\lambda-\frac{2}{3}g'B_m)C'/4\\
\zeta-\bar{\zeta}~{\rm sector},~G_{\mu\bar{\nu}}&=&l^2 {\delta_\mu}^\nu C'/2.
\end{eqnarray}
Above, $W$ refers to the left weak boson field (with coupling constant $g$),
 $B$ to its hypercharge counterpart (with coupling constant $g'$/2),
$V$ to the gluon field (with coupling constant $f$), $\lambda$ to the SU(3) lambda matrices;
$A$ is a generic full vector set comprising all those components 
multiplied by their coupling constants.
$C$ and $C'$ are in principle independent curvature polynomials involving the various
invariants $\bar{\zeta}\zeta$ associated with the gauge subgroups. To simplify the argument
we will take $C=C'$ to be universal for gauge fields and gravity as this
has little effect on the conclusions. For the standard model this means that $C$ can be a
direct product of colour, left and right permissible polynomial factors; thus
\begin{equation}
C=(1+c_1Z+c_2Z^2+c_3Z^3)(1+c_RZ_R+c_{RR}Z_R^2)(1+c_LZ_L+c_{LL}Z_L^2)
\end{equation}
introduces seven independent curvature coefficients $c_i$. The Berezinian is easily
evaluated as a product of the three subgroups (trivial for SU(2)), viz.
\[
\sqrt{{\rm sdet}\,G..}=\sqrt{g..}[1-c_1Z+(c_1^2-c_2)Z^2-(c_1^3-2c_1c_2+c_3)Z^3].
\]

Referring to Table 1 in reference [2], we may extract the total gauge field contributions to be
\begin{equation}
\frac{1}{3!2!2!}\left(\frac{l^2}{2}\right)^6\!\!\!\int\!\!d^3\zeta\ldots d^2\zeta_L \sqrt{-G..}{\cal R}
= \sqrt{-g..}\left(\frac{{\cal A}R^ {[g]} }{l^2} + {\cal B}{\rm Tr}(F.F) + \frac{{\cal C}}{l^4}\right), 
\end{equation}
\begin{equation}
{\rm where~~~~}{\cal A}=4(2c_1^3-3c_1c_2+c_3)(c_{RR}-c_R^2)(c_{LL}-c_L^2),
\end{equation}
\begin{eqnarray}
{\cal B}{\rm Tr}(F.F)&=&\frac{1}{12}(3c_1^2-2c_2)(c_{LL}-c_L^2)(c_{RR}-c_R^2)
                 [f^2V^{mn}.V_{mn} +2g'^2B^{mn}B_{mn}/3]\nonumber \\
  & &-\frac{1}{2}(2c_1^3-3c_1c_2+c_3)c_R (c_{LL}-c_L^2)\,g'^2B^{mn}B_{mn}\nonumber\\
   & &\!\!\!\!\!-\frac{1}{4}(2c_1^3-3c_1c_2+c_3)c_L (c_{RR}\!-c_R^2)
            [g^2W^{mn}.W_{mn} + g'^2 B^{mn}B_{mn}]
\end{eqnarray}
\begin{eqnarray}
-{\cal C} &=&48(5c_1^4-10c_1^2c_2+2c_2^2+3c_1c_3)(c_{RR}-c_R^2)(c_{LL}-c_L^2)+
\nonumber\\
&& 8c_L(c_{RR}-c_R^2)(4c_{LL}-3c_L^2)(2c_1^3-3c_1c_2+c_3)+\nonumber\\
&& 8c_R(c_{LL}-c_L^2)(4c_{RR}-3c_R^2)(2c_1^3-3c_1c_2+c_3).
\end{eqnarray}
As perusual, $V_{mn}\equiv V_{n,m}-V_{m,n}+if[V_m,V_n]$, 
$W_{mn}\equiv W_{n,m}-W_{m,n}+ig[W_m,W_n]$ and
$B_{mn}\equiv B_{n,m}-B_{m,n}$ retain their significance of `curls' or field strengths
of the three gauge fields in question.

Because $C$ multiplies the gravitational field and we believe that gravity is parity invariant
classically, we can with some confidence assume that $c_R=c_L, c_{RR}=c_{LL}$; 
this serves to simplify the above 
expressions even more and reduce the number of curvature coefficients to five. 
Much more importantly, we need to ensure that the normalization is
identical for all the gauge fields, otherwise the stress tensor will not couple universally
to all sources, as it must. Focussing on the gluon and weak boson fields, this means that
$f^2(2c_2-3c_1^2)(c_{LL}-c_L^2)=3g^2c_L(2c_1^3-3c_1c_2+c_3)$. If we then apply
the same condition for the hypercharge field, we deduce that
\[
g^2 = 3g'^2 + 2g^2g'^2/3f^2,
\]
which is a relation between the three coupling constants.  A more meaningful version
is found by remembering that $g'=g\tan\theta_w,\, e=g\sin\theta_w$, where 
$\theta_w$ is the weak mixing angle. Hence
\begin{equation}
1= 3\tan^2\theta_w +2e^2\sec^2\theta_w/3f^2~~{\rm or}~~ 4\sin^2\theta_w=1-2e^2/3f^2
=1-2\alpha/3\alpha_s.
\end{equation}
Strong chromic properties are therefore responsible for diminishing the weak angle from
the purely leptonic result of $\tan^2\theta_w=1/3$. Since the ratio $\alpha/\alpha_s$ runs, 
according to the renormalization group, let us estimate this reduction by choosing a weak scale
of about 100 GeV when $\alpha\simeq 0.008$ and $\alpha_s\simeq 0.115$; we then
obtain $\sin^2\theta_w \simeq 0.238$. Bearing in mind that we have not considered
quantum corrections,  the result is not far off the standard experimental
estimates\cite{Moller,LowEnergy,DLM} hovering around 0.232.

Extracting a common factor $g^2c_L(c_{LL}-c_L^2)(2c_1^3-3c_1c_2+c_3)/2$ 
from (32)-(34) we get
\begin{eqnarray}
\int\!\!d^3\zeta\ldots d^2\zeta_L \sqrt{-G..}{\cal R}/\sqrt{-g..}&=&
\frac{4(c_{LL}-c_L^2)}{l^2g^2c_L}R^ {[g]}- \nonumber \\
&&\,\,(V^{mn}.V_{mn}+W^{mn}.W_{mn}+B^{mn}B_{mn})/4\nonumber\\
&&\hspace{-3.5cm}-\frac{16}{g^2l^4}\left[
\frac{3(c_L^2-c_{LL})(5c_1^4-10c_1^2c_2+2c_2^2+3c_1c_3)}{c_L(2c_1^3-3c_1c_2+c_3)}
+(3c_L^3-4c_{LL})\right]
\end{eqnarray}
whence we may identify the Newtonian constant
\begin{equation}
64\pi G_N=l^2g^2c_L/(c_{LL}-c_L^2)
\end{equation}
and the cosmological constant $\Lambda$ via
\begin{equation}
\frac{\Lambda}{8\pi G_N} = \frac{16}{l^4g^2}\left[
\frac{3(c_L^2-c_{LL})(5c_1^4-10c_1^2c_2+2c_2^2+3c_1c_3)}{c_L(2c_1^3-3c_1c_2+c_3)}
+(3c_L^2-4c_{LL})\right]
\end{equation}
So far we have not succeeded in establishing a principle for restricting property curvature 
coefficients, and can only deduce that the $c$'s are constrained by the conditions,
\[
c_L\neq 0,\quad c_{LL}\neq c_L^2,\quad  c_L/(c_{LL}-c_L^2)>0
\]
and the bracketted expression in (38) is positive. The observational smallness of $\Lambda$ indicates that both\footnote{One can only make guesses as to what values the curvature 
coefficients take. 
With no principle to guide us this disappointment is mitigated by the freedom afforded us by the $c_i$ in arranging the cosmological constant to fit observation.}
\[ 
4c_{LL}\simeq 3c_L^2 {\rm ~and~} 5c_1^4-10c_1^2c_2+2c_2^2+3c_1c_3\simeq 0.
\]

We still have to describe the interactions of the gauge fields with the matter fields annotated
in section 3. The procedure is rather automatic and hails from the frame vectors:
\begin{eqnarray*}
2{E_a}^\zeta &=&i\!\left(\!fV_a.\lambda\!-\!\frac{2}{3}g'B_a\!\right)\!\zeta,\,
2{E_a}^{\zeta_L}\!\!=i(gW_a.\tau\!-\!g'B_a)\zeta_L,\,
2{E_a}^{\zeta_R}\!\!=ig'B_a(\tau_3\!-\!1)\zeta_R\\
2i{E_a}^{\bar{\zeta}}&\!=&\bar{\zeta}\!\left(\!fV_a.\lambda\!-\!\frac{2}{3}g'B_a\!\right)\!,
2i{E_a}^{{\bar\zeta}_L}\!\!=\bar{\zeta}_L(gW_a.\tau\!-\!g'B_a),
2i{E_a}^{\bar{\zeta}_R}\!\!=\bar{\zeta}_Rg'B_a(\tau_3-1),
\end{eqnarray*}
which are the primary source of the supermetric elements (25)-(28). These frame vectors
are, in this scheme, associated with the covariant derivative
\begin{eqnarray}
D_a = {E_a}^M\partial_M&=&{E_a}^m\partial_m+{E_a}^\zeta\partial_{\zeta} 
         +{E_a}^{\bar{\zeta}}\partial_{\bar{\zeta}}+\nonumber\\
         &&{E_a}^{\zeta_L}\partial_{\zeta_L}+{E_a}^{\bar{\zeta}_L}\partial_{\bar{\zeta}_L}+
         {E_a}^{\zeta_R}\partial_{\zeta_R}+{E_a}^{\bar{\zeta}_R}\partial_{\bar{\zeta}_R}.
\end{eqnarray}

To see how this automatically produces the interactions of gauge fields with source fields,
we shall simplify the presentation to start with by going to flat space, pretending there is no 
property curvature nor any spacetime curvature. Thus assume a Minkowskian 
background and a trivial curvature polynomial ($C$=1 or $c_i\!=\!0$). 
Consider first the four generations
of leptons ${\cal L}, {\cal L}', {\cal L}'', {\cal L}'''$ where the generic doublet is
${\cal L}_{\{L,R\}}=(\nu_{\{L,R\}},\ell_{\{L,R\}})$. Ignoring
conjugate contributions which only serve to double the results, these leptonic doublets
arise in the selfdual superfield $\Psi$ combination as follows:
\begin{eqnarray}
2\Psi &\supset& \,\,\,[\bar{\zeta}_L{\cal L}_L(1-Z^3/6)+\bar{\zeta}_L{\cal L}'_L Z(1-Z/2)/\sqrt{3}]
Z_R(1-Z_L)\nonumber\\
&&+\,[\bar{\zeta}_L{\cal L}_L''(1-Z^3/6)+\bar{\zeta}_L{\cal L}'''_L Z(1-Z/2)/\sqrt{3}]
(1-Z_R^2/2)(1-Z_L),
\nonumber\\
&&\,\,\,+ (L\leftrightarrow R)
\end{eqnarray}
\begin{eqnarray}
2\overline{\Psi} &\supset& -\,[\overline{{\cal L}}_L\zeta_L(1-Z^3/6)
+\overline{{\cal L}}'_L \zeta_LZ(1-Z/2)/\sqrt{3}]
Z_R(1-Z_L)\nonumber\\
&&+\,[\overline{{\cal L}}_L''\zeta_L(1-Z^3/6)+\overline{{\cal L}}'''_L\zeta_L Z(1-Z/2)/\sqrt{3}]
(1-Z_R^2/2)(1-Z_L),
\nonumber \\
&&\,\,\,+ (L\leftrightarrow R).
\end{eqnarray}
In this flat limit one readily checks that
\[
\int\!\!d^3\zeta\ldots d^2\zeta_L\,\overline{\Psi}\gamma.\partial\Psi=
   \overline{{\cal L}}\gamma.\partial{\cal L} +  \overline{{\cal L}}'\gamma.\partial{\cal L}'
   + \overline{{\cal L}}''\gamma.\partial{\cal L}'' +  \overline{{\cal L}}'''\gamma.\partial{\cal L}'''.
\]
Moreover a mass term $\bar{\Psi}\Psi$ vanishes upon integration for two reasons: 
chiral orthogonality
and a mismatch between powers of left and right properties; this indicates introduction
of a Bose field which straddles chirality and cures the property mismatch too.

Upon including the frame vectors as in (39), introducing spacetime
curvature and noting that the leptonic components in (40) only involve the left/right
properties, we see that
\begin{equation}
i\gamma^aD_a\Psi \supset i\gamma^a(\partial_a\!+\!{E_a}^{\bar{\zeta}_L}\partial_{\bar{\zeta}_L})
    \bar{\zeta}_L{\cal L} +..= \bar{\zeta}_L\gamma^a{e_a}^m\left[
             i\partial_m\!-\!\frac{1}{2}(gW_m.\tau\!-\!g'B_m)\right]{\cal L} + ..
\end{equation}
and similarly for the right parts. We thereby recover the standard gauge interactions for
every leptonic generation.

With quarks the problem is slightly more complicated in that we must include the chromic
properties; to illustrate what happens, consider the first two generations and
ignore curvature (where ${\cal Q}\zeta\zeta$ signifies ${\cal Q}^1\zeta^2\zeta^3$ plus perms):
\begin{eqnarray}
4\Psi &\supset& [(\zeta_L^{\bar{0}}U_L+\zeta_L^{\bar{4}}D_L)Z_R
           +(\zeta_L^{\bar{0}}U'_L+\zeta_L^{\bar{4}}D'_L)(1-Z_R^2/2)](1+Z_L)\zeta\zeta(1-Z)\\
        &&+ (L\leftrightarrow R)  + \ldots \nonumber
\end{eqnarray}
This is correctly normalized in as much as 
\[
\int\!\!d^3\zeta\ldots d^2\zeta_L\,\overline{\Psi}\gamma.\partial\Psi
= (\bar{U}\gamma.\partial U+ \bar{D}\gamma.\partial D)
  +(\bar{U}'\gamma.\partial U' + \bar{D}'\gamma.\partial D') + \ldots
\]
Noting that the chromic product $\zeta\zeta$ behaves like $\bar{\zeta}$ in (43), 
we see that the inclusion of the frame vectors  from (39) brings in 
\begin{equation}
2D_a\!=\!e_a^m\!\!\left[2\partial_m\!+\!i(fV_m.\lambda\!-\!\frac{2}{3}B_m)\zeta
\partial_{\zeta}\!-\! i\bar{\zeta}_L(gW_m.\tau\!-\!g'B_m)\partial_{\bar{\zeta}_L}
\!\!-\!i g'B_m(\tau_3\!-\!1)\bar{\zeta}_R\partial_{\bar{\zeta}_R}\!\right]
\end{equation}
which yields precisely the correct interactions with gluons and electroweak 
gauge bosons of each quark generation.

It only remains to describe what effect the property curvature polynomial $C$ in (30)
has on these standard results, where we previously set $C=1$. For example
return to the first two quark (${\cal Q}=(U,D)$) generations of the superfield $\Psi$ in (40)
to discover the effect. Because ${E_a}^M$ contains
the factor $1/\sqrt{C}$ as well the overall factor $\sqrt{-G..}$, this will induce a mixing 
and renormalization of the quark fields due to the various curvature coefficients. From
(43) we get,
\[
\Psi\propto \bar{\zeta}_L[{\cal Q}Z_R+{\cal Q}'_L (1-Z_R^2/2)](1+Z_L)\zeta\zeta(1-Z)
 +(L\leftrightarrow R),
\]
and the factorised expression
\begin{eqnarray*}
\frac{\sqrt{-G..}}{\sqrt{C}}&=&\sqrt{-g..}\left[1-\frac{3}{2}c_1Z+..\right]
   \left[1-\frac{1}{2}c_LZ_L+\frac{1}{2}(\frac{3}{4}c_L^2-c_{LL})Z_L^2\right]\\
 && \qquad\qquad\qquad\qquad\quad\times\left[1-\frac{1}{2}c_RZ_R+
     \frac{1}{2}(\frac{3}{4}c_R^2-c_{RR})Z_R^2\right].
\end{eqnarray*}
Imposing the parity even conditions $c_R=c_L, c_{RR}=c_{LL}$, 
we deduce that the quark currents are the mixtures:
\[
(1+c_L/4)(1+3c_1/2)[\bar{\cal Q}\gamma{\cal Q}+x(\bar{\cal Q}'\gamma{\cal Q}
 + \bar{\cal Q}\gamma{\cal Q}') + y\bar{\cal Q}'\gamma{\cal Q}'],
\]
where $x=-c_L/2, \, y= 1+c_{LL}/2-3c_L^2/8$. These currents can be diagonalised by 
taking the combinations ${\cal Q}+x{\cal Q}', \, {\cal Q}'$ and carrying out further
wave renormalizations on each of the mixed fields. It should be emphasized that
these mixings and rescalings affect the kinetic energy of the quarks and their
gauge interactions {\em equally} so no further scalings of coupling constants are needed.
The very same phenomenon extends to the other quark generations as well as
to leptons.

\subsection{Yukawa interactions}
Our next port of call is the Bose superfield $\Phi$ since it connects left to right fermion sectors.
As will be seen from (21) and (22) there are numerous bosonic colour
triplets and octets; but on the presumption that colour is always confined let us concentrate
on the colour singlets; these contain the Higgs fields whose expectation values are
adduced to give rise to the fermion and gauge boson masses. Consider then a hermitian
superfield in flat space,
\begin{eqnarray}
\Phi &\supset& (1+Z^3/6)[\bar{\zeta}_L\varphi\zeta_R(1+Z_R)(1+Z_L)+
       \bar{\zeta}_L\varphi'\zeta_R(1-Z_R)(1-Z_L)\nonumber \\
       & & \qquad\qquad\quad+\bar{\zeta}_R\varphi^\dag\zeta_L(1+Z_L)(1+Z_R)
       + \bar{\zeta}_R\varphi'^\dag\zeta_L(1-Z_L)(1-Z_R)]-\nonumber \\
       & & (1/\sqrt{3})Z(1+Z/2)[\bar{\zeta}_L\varphi''\zeta_R(1+Z_R)(1+Z_L)+
      \bar{\zeta}_L\varphi'''\zeta_R(1-Z_R)(1-Z_L)\nonumber \\
       & & \qquad\qquad\quad+\bar{\zeta}_R\varphi''^\dag\zeta_L(1+Z_L)(1+Z_R)
      + \bar{\zeta}_R\varphi'''^\dag\zeta_L(1-Z_L)(1-Z_R)].
\end{eqnarray}
Imposing parity evenness under $L\leftrightarrow R$, so $\varphi=\varphi^\dag$, etc.
we may reduce (45) to the components
\begin{eqnarray}
\Phi &\supset& (1+Z^3/6)[\bar{\zeta_L}\varphi\zeta_R(1+Z_R)(1+Z_L)+
       \bar{\zeta}_L\varphi'\zeta_R](1-Z_R)(1-Z_L)] -\nonumber \\
       & & Z(1\!+\!Z/2)[\bar{\zeta_L}\varphi''\zeta_R(1\!+\!Z_R)(1\!+\!Z_L)+
       \bar{\zeta}_L\varphi'''\zeta_R](1\!-\!Z_R)(1\!-\!Z_L)]/\sqrt{3},
\end{eqnarray}
plus terms where $R\leftrightarrow L$. The normalization of the 4 possible real Higgs quartets, 
$\varphi=(\phi_0I+\tau.\phi)/\sqrt{2}$,  then emerges correctly:
\[
\int\!\!d^3\zeta\ldots d^2\zeta_L\,\, \Phi^2 \propto 
{\rm Tr}\,[\varphi^2 +  \varphi'^2 + \varphi''^2 + \varphi'''^2] + \ldots
\]
Each of these four $\varphi$ fields couples in the same way to gauge fields when one works
out the the action $D_a\Phi\,D^a\Phi$ because the covariant derivative $D$, as in (39),
acts in the expected manner on the properties $\zeta_L, \zeta_R$ and the $Z$ are
blind to its action. Previously, when we were dealing with the purely leptonic case, 
we had just one Higgs quartet and
effectively one interaction with gauge bosons. We must now confront four possible ones,
but all of the same type. At this point, without a full investigation of the potential 
$V(\Phi)$ responsible for producing classical expectation values of 
$\langle\varphi\rangle$, we only know that a vacuum state arises through a 
linear combination of each of the four generic\cite{RD} terms $\langle \phi_0+\phi_3\rangle$
and we cannot say much more. In any event, when one turns on spacetime and
property curvature, via ${e_a}^m(x)$ and $C(Z)$, the ensuing fields will conform
to general relativity and states with the same quantum numbers will mix (and require
rescaling) depending on the magnitudes of the curvature coefficients $c_i$. This
mirrors the fermions.

The Yukawa interactions resemble those of the standard model but are yet different.
Thus the standard model in its simplest form entertains a single Higgs field $\phi$ with
independent couplings ${\frak g}_i$ to each of the sources $\psi_i$; in the present scheme
we encounter a single Yukawa superfield\footnote{Below we have ignored the duality
choices mentioned at the start of section 3.2 since these do not couple to the
fermion superfield $\Psi$ upon integrating over $\zeta_L$ and $\zeta_R$,
at least in the flat limit.} coupling ${\frak g}\overline{\Psi}\Phi\Psi$ 
where $\Phi$ encompasses the several Higgs fields $\varphi_i$. To see how this 
works out in practice, study the interactions of the component fields in (46) say with
the source fields for leptons  and quarks as in (40 and (43). The calculations are a bit messy 
but straightforward: one simply has to collect the correct powers of property before
integrating over them. In the quark sector, with reference to (43) and (46), we obtain
such interactions as
\begin{eqnarray}
\int\!\!d^3\zeta\ldots d^2\zeta_L\,\, \Psi\Phi\Psi  &\propto& 
(\bar{\cal Q}_R+2\bar{\cal Q}'_R)(\varphi-\varphi''/\sqrt{3})({\cal Q}_L+2\cal{Q}'_L)\nonumber\\
 & & + \bar{\cal Q}_R(\varphi' -\varphi'''/\sqrt{3}){\cal Q}_L + (L\leftrightarrow R).
\end{eqnarray}
For leptons we get a similar result:
\begin{eqnarray}
\int\!\!d^3\zeta\ldots d^2\zeta_L\,\, \Psi\Phi\Psi  &\propto& 
 \bar{\cal L}\varphi{\cal L}+(\bar{\cal L}+2\bar{\cal L}'')\varphi'({\cal L}+2{\cal L}'')
    \nonumber \\
&& + 2\bar{\cal L}'\varphi{\cal L}' +2(\bar{\cal L}'+2\bar{\cal L}''')\varphi'
({\cal L}'+2{\cal L}''') \nonumber \\
&& +2\left[ \bar{\cal L}'\varphi''{\cal L}' +
(\bar{\cal L}'+2\bar{\cal L}''')\varphi'''({\cal L}'+2{\cal L}''') \right]/\sqrt{3}.
\end{eqnarray}
Introducing curvature of spacetime through $\det{e}$ renders the answers generally
covariant, while inclusion of property curvature $C(Z)$ causes the various leptonic and
quark components to mix even more through the curvature coefficients $c_i$, 
as we saw for fermions, but {\em without disturbing the gauge couplings} The whole
edifice becomes rather complicated and warrants future computational analysis though
it is conceptually simple.

\section{Concluding Remarks}
This work represents the culmination of a series of articles exploring the picture that the 
``extra dimensions'' which everyone is seeking to append to spacetime are simply 
the properties of a system, represented by scalar anticommuting complex variables. In
earlier articles we investigated what happens when we added the properties successively
and we gradually gained an understanding of how the scheme operates; with this
knowledge we have in this paper incorporated the full gamut characterising the SM: 
chromicity, left and right electricity/neutrinicity.  In this scenario an event---where something necessarily  changes---is fully described by where, when, and {\em what} happens.

A consequence of this approach is that gravity and the gauge interactions of the standard
model are fully unified in a generalised metric containing curvature of spacetime as well
as property. The ensuing field theory and particle spectrum demands the imposition of Grassmannian duality, as explained in sections 2 and 3. Even if our choices of dualities 
within the subgroups of the full group turn out to be awry, the need for some kind of duality 
is imperative and emergence of generations is inevitable. 
However, assuming our choices are the correct ones, we were able to 
accommodate the known particle generations, but in the process we were led
to predict other generations of quarks and leptons, and especially the existence of
a quark carrying charge -4/3, accompanying $t~ \&~ b$ say in a weak isospin triplet.
Another feature was the emergence of totally sterile scalar states (8 of them), singlets of
all the gauge groups and therefore decoupled from standard matter; these may
or may not have some connection with dark matter---it is too early to say.
The marriage of gravity to the gauge fields via the frame vectors and insistence on the 
universal character of coupling to gravity resulted in the
prediction of a weak mixing angle which is close to experiment.

Apart from researching the mechanism generating masses through expectation values of 
at least four standard Higgs fields which couple to the sources, the property curvature 
coefficients stick out as the only vague feature of this scheme. We have been unable
to formulate a principle which determines the $c_i$, which in turn fix the 
cosmological constant, etc. What we do know is that pure property transformations
like $\zeta \rightarrow \zeta\,f(\bar{\zeta}\zeta)$, which do not affect gauge
variations, produce nasty looking off-diagonal elements in the property-property
sector via the altered metric $d\bar{\zeta}\zeta\zeta d\bar{\zeta}$, etc. Conversely
those property transformations can be used to eliminate such unpleasant 
contributions, but will not fix the curvature coefficients. This is the most pressing
problem confronting us.  Only after its proper resolution might we turn to
the BRST quantization since there seems to be a natural place for it within the current 
framework. A separate issue is the calculation of masses and mixings, where $c_i$ enter
once again. In our opinion, It seems unlikely that we will be able to cover the 
huge and diverse range of masses (from neutrinos to the top quark and mixings) 
with this set of $c_i$ and various $\langle\varphi\rangle$; most probably some dynamical
mechanism will be needed to produce the masses of the lighter leptons through their
weak interactions and the $c_i$ may themselves be expectation values of additional
scalar fields.

If right gauge bosons are experimentally discovered, they are readily incorporated by 
introducing additional frame vectors affecting the right properties $\zeta_R$. In the end 
like everyone else we 
are hostage to experiment and to what the LHC and future colliders 
will reveal. It is entirely possible that this whole scheme will fail miserably by not agreeing with
observations and thus turn out to be a figment of imagination. Hopefully some of its ingredients may
survive but only time will tell.

\section*{Acknowledgements}

Brian Kenny has been urging me over the years to take the `next step' in 
this approach to unification. I thank him for his keen interest and continued faith in this 
program.

\newpage
\appendix
\section{Rule for $N$ properties integration}

In the text we are confronted with integration over properties $\zeta,\bar{\zeta}$. 
Our convention for handling this is as follows.  The Grassmannian measure is to be taken
in the order 
$(d\zeta^1d\zeta^{\bar{1}})(d\zeta^2d\zeta^{\bar{2}})\ldots (d\zeta^Nd\zeta^{\bar{N}})  =
(-1)^{<N/2>}(d\zeta^1d\zeta^2\ldots d\zeta^N)
(d\zeta^{\bar{1}}d\zeta^{\bar{2}}\ldots d\zeta^{\bar{N}}),$
where $<N/2>$ signifies the integer part of $N/2$.
We define invariants $Z\equiv \zeta^{\bar{\mu}}\zeta^\mu$ for each of the subgroups 
(with appropriate labelling on $Z$) such that for SU($N$),
\[
\int (d\zeta^1d\zeta^{\bar{1}})(d\zeta^2d\zeta^{\bar{2}})\ldots (d\zeta^Nd\zeta^{\bar{N}})
\,\,Z^N = N!
\]

\section{Rule for SU($N$) duality}
If one substitutes $(r,s)\rightarrow (N-s,N-r)$ in the expansion of a superfield in powers
$(\zeta)^r(\bar{\zeta})^s$ it is a fact that the associated SU($N$) representations are duplicated, corresponding to reflection about the cross-diagonal We call the act of cross-diagonal 
reflection the `duality' operation and indicate it by the symbol $\times$. 
Specifically our rule for determining the factors arising from this reflection is given by
\[
(\zeta^{\bar{\mu}_1}..\zeta^{\bar{\mu}_r}\zeta^{\nu_1}..\zeta^{\nu_s})^\times
\!\!=\!(\epsilon^{\bar{\mu}_1.\bar{\mu}_r\rho_1.\rho_{N\!-\!r}}\zeta^{\rho_{N\!-\!r}}..\zeta^{\rho_1})
(\zeta^{\bar{\sigma}_{N\!-\!s}}..\zeta^{\bar{\sigma}_1}
\epsilon^{\sigma_1.\sigma_{N\!-\!s}\nu_1.\nu_s})/(N-r)!(N-s)!
\]
This looks rather complicated, but three examples may make this rule clearer. For instance
consider colour ($N=3$); then the dual of a typical octet element is associated with
a - sign as follows:
\[
(\zeta^{\bar{1}}\zeta^2)^\times = \epsilon^{\bar{1}\bar{2}\bar{3}}\zeta^3\zeta^2\,
\zeta^{\bar{1}}\zeta^{\bar{3}}\epsilon^{312} = -\zeta^{\bar{1}}\,\zeta^2\zeta^{\bar{3}}\zeta^3
= -\zeta^{\bar{1}}\zeta^2\,Z,
\]
whereas for the singlet product of the three colours there is no sign change:
\[
(\zeta^{\bar{1}}\zeta^{\bar{2}}\zeta^{\bar{3}})^\times = 
\epsilon^{\bar{1}\bar{2}\bar{3}}\,\zeta^{\bar{3}}\zeta^{\bar{2}}\zeta^{\bar{1}}\epsilon^{123}
= \zeta^{\bar{1}}\zeta^{\bar{2}}\zeta^{\bar{3}}.
\]
Similarly for the triplet,
\[
(\zeta^{\bar{1}}Z)^\times = (\zeta^{\bar{1}}(\zeta^{\bar{2}}\zeta^2+\zeta^{\bar{3}}\zeta^3))^\times
= \epsilon^{\bar{1}\bar{2}\bar{3}}\zeta^3 \zeta^{\bar{3}}\zeta^{\bar{1}}\epsilon^{132} +
  \epsilon^{\bar{1}\bar{3}\bar{2}}\zeta^2 \zeta^{\bar{1}}\zeta^{\bar{2}}\epsilon^{213}
  = -\zeta^{\bar{1}}Z.
\]
By these sorts of manipulations we can ascertain the duality signs for all the entries in
the expansion table, as stated in the text; applying selfduality or anti-selfduality can 
the sometimes serve to eliminate some of the representations sitting on the cross-diagonal.
Finally, it goes without saying that dualizing twice results in the identity operation.

\end{document}